\documentclass[aps,pra,twocolumn,groupedaddress]{revtex4}

\usepackage{graphicx}
\usepackage{amsmath}
\usepackage{amssymb}
\usepackage{amsthm}
\theoremstyle{definition} \newtheorem{Defi}{Definition}
\theoremstyle{plain} \newtheorem{theorem}{Theorem}

\begin{document}


\title{An efficient quantum search engine on unsorted database}


\author{Heping Hu}
\email[]{e-mail: newpine2002@yahoo.com.cn}
\author{Yingyu Zhang}
\author{Zhengding Lu}
%

\affiliation{Huazhong University of Science and Technology}


\date{\today}

\begin{abstract}
We consider the problem of finding one or more desired items out of
an unsorted database. Patel has shown that if the database permits
quantum queries, then mere digitization is sufficient for efficient
search for one desired item. The algorithm, called factorized
quantum search algorithm, presented by him can locate the desired
item in an unsorted database using $O(log_{4}N)$ queries to
factorized oracles. But the algorithm requires that all the property
values must be distinct from each other. In this paper, we discuss
how to make a database satisfy the requirements, and present a
quantum search engine based on the algorithm. Our goal is achieved
by introducing auxiliary files for the property values that are not
distinct, and converting every complex query request into a sequence
of calls to factorized quantum search algorithm. The query
complexity of our algorithm is $O(P*Q*M*log_{4}N)$, where P is the
number of the potential simple query requests in the complex query
request, Q is the maximum number of calls to the factorized quantum
search algorithm of the simple queries, M is the number of the
auxiliary files for the property on which our algorithm are
searching for desired items. This implies that to manage an unsorted
database on an actual quantum computer is possible and efficient.
\end{abstract}


\maketitle

\section{Introduction}
Consider a collection of items in a database, characterized by some
property with values on the real line. Sorting these items means
arranging them in an ordered sequence according to their property
values. The purpose of sorting is to facilitate subsequent searches
of that database for items with known values of the property
\cite{Patel01}.

The best classical algorithm uses $log_{2}N$ queries to find a
desired item out of an ordered list of N items, while a quantum
computer can solve the problem using a constant factor fewer
queries. The best known lower bound shows that any quantum algorithm
for the problem requires at least $(lnN-1)/\pi\approx 0.221log_{2}N$
queries \cite{Hoyer00}. The algorithm obtained by Farhi, Goldstone,
Gutmann, and Sipser, uses $3\lceil log_{52} N\rceil$ queries,
showing that a constant factor speedup is indeed possible
\cite{Farhi99}. The best published exact quantum algorithm for the
problem uses $4log_{605}N\approx 0.433log_{2}N$ queries
\cite{Childs07}.

In this paper, we consider the problem of finding one or more
desired items out an unsorted database. Patel has shown that if the
database quantum queries, then mere digitization is sufficient for
efficient search for one desired item. The algorithm, called
factorized quantum search algorithm, presented by him can locate the
desired item in an unsorted database using $O(log_{4}N)$ queries to
facorized oracles \cite{Patel01}. Unfortunately, the algorithm would
not work if it is used to search more than one desired items out of
an unsorted database, or the property values among which it is
searching for the desired item are not distinct. From the point of
view of unsorted database manipulation, we consider the problem with
more than one desired items in an unsorted database and with the
property values that are not distinct from each other. And we
present a quantum search engine, based on the factorized quantum
search algorithm, that can solve the problems. Our goal is achieved
by introducing auxiliary files for the property values that are not
distinct, and converting the searching request into a sequence of
calls to factorized quantum search algorithm.

The paper is organized as follows: In section II, we review quantum
random access memory, Structured Query Language, and factorized
quantum search algorithm. In section III, we present our quantum
search engine, discuss how to digitize the property values of an
unsorted database, which might be distinct from each other or not,
and discuss how to convert a complex query request into a sequence
of calls to the factorized quantum search algorithm. We also discuss
how the factorized quantum search algorithm works on an actual
quantum computer with the unsorted database existing in the quantum
random access memory. In section IV, We give several examples to
show how the quantum search engine works. We conclude the paper in
section V.

\section{Backgrounds}

\subsection{Quantum random access memory} A quantum random access
memory (qRAM) uses nqubits to address any quantum superposition of N
memory cells. It means the qRAM can perform memory accesses in
coherent quantum superposition \cite{Nielson00}: if the quantum
computer needs to access a superposition of memory cells, the
address register A must contain a superposition of addresses
$\sum_{j}a_{j}|j\rangle_{A}$, and the qRAM will return a
superposition of data in a data register D, correlated with the
address register:
\begin{equation}
\sum_{j}a_{j}|j\rangle_{A} \longrightarrow
\sum_{j}a_{j}|j\rangle_{A}|d_{j}\rangle_{D},
\end{equation}
where $d_{j}$ is the content of the jth memory cell
\cite{Giovannetti08}. Giovannetti, Lloyd and Maccone have presented
an architecture that exponentially reduces the requirements for a
memory call: $O(log_{2}N)$ switches need be thrown instead of the N
used in classical RAM designs, which yields a more robust qRAM
algorithm than conventional (classical and quantum) designs. See
\cite{Giovannetti08} for more details.

\subsection{Structured query language} Structured Query Language (SQL)
is a tool widely used in manipulating the classical databases
\cite{Elmasri06}.Basic operations in SQL include inserting a new
record to a database file (INSERT), updating an existing record
(UPDATE), deleting an exiting record (DELETE), selecting (SELECT)
and performing an arbitrary operation on some records, backing up a
portion of a database (BACKUP). and restoring the backup (RESTORE)
\cite{Younes07}. The most commonly used operation among them might
be the SELECT operation, which is defined as:

\begin{picture}(180,160)
\begin{footnotesize}
\put(0,145) {SELECT [ ALL $|$ DISTINCT [ ON ( expression[, ...])]]}

\put(0,135){* $|$ expression [ AS output\_name ] [, ...]}

\put(0,125){[ INTO [ TEMPORARY $|$ TEMP ] [ TABLE ] new\_table ]}

\put(0,115){[ FROM from\_item [, ...] ]}

\put(0,105){[WHERE condition ]}

\put(0,95){[ GROUP BY expression [, ...] ]}

\put(0,85){[ HAVING condition [, ...] ]}

\put(0,75){[ {UNION $|$ INTERSECT $|$ EXCEPT [ ALL ] } select ]}

\put(0,65){[ ORDER BY expression [ ASC $|$ DESC $|$}

\put(0,55){USING operator ] [, ...] ]}

\put(0,45){[FOR UPDATE [ OF class\_name [, ...] ] ]}

\put(0,35){[ LIMIT { count $|$ ALL } [ { OFFSET $|$ , } start ]]}
\end{footnotesize}

\put(10,20){FIG. Definition of the SELECT operation}
\end{picture}

The SELECT statement is used to form queries for extracting
information out of the database. The details about how to use it and
the other basic operations can be found in \cite{Elmasri06}. We only
show several examples, about the SELECT operation, useful for our
discussion below. Assume that we have such a table as follows stored
in the database, and the name of the table is Table1.

\begin{picture}(120,110)
\put(40,60){
\begin{tabular}{|c |c |c |}\hline
ID & Name & Age\\\hline

960112 & Lin & 18 \\\hline

960114 & Yang & 20 \\\hline

960113 & Xing & 19 \\\hline

960115 & Yingyu & 20 \\\hline

\end{tabular}
}

\put(60,15){Table. 1}

\end{picture}

Example 1:

\begin{picture}(100,40)
\begin{footnotesize}
\put(30,30){SELECT ID, Name, Age}

\put(30,20){FROM Table1}

\put(30,10){WHERE Age=20}
\end{footnotesize}
\end{picture}

It means selecting the records whose property values of Age is 20.
This operation will return with

\begin{picture}(120,80)
\put(40,40){
\begin{tabular}{|c |c |c|}\hline
ID & Name & Age\\\hline

960114 & Yang & 20\\\hline

960115 & Yingyu & 20\\\hline

\end{tabular}
}

\put(20,10){Table. 2: Result of example 1.}

\end{picture}

Example 2:

\begin{picture}(100,40)
\begin{footnotesize}
\put(30,30){SELECT ID, Name, Age}

\put(30,20){FROM Table1}

\put(30,10){WHERE ID$<$960115}
\end{footnotesize}
\end{picture}

Similarly, this operation will return with

\begin{picture}(120,90)
\put(40,50){
\begin{tabular}{|c |c |c |}\hline
ID & Name & Age\\\hline

960112 & Lin & 18 \\\hline

960114 & Yang & 20 \\\hline

960113 & Yang & 19 \\\hline

\end{tabular}
}

\put(20,10){Table. 3: Result of example 2.}

\end{picture}

Example 3:

\begin{picture}(100,40)
\begin{footnotesize}
\put(30,30){SELECT ID, Name, Age}

\put(30,20){FROM Table1}

\put(30,10){WHERE ID$<$960115 AND Age=20}
\end{footnotesize}
\end{picture}

It will return with

\begin{picture}(120,65)
\put(40,35){
\begin{tabular}{|c |c |c |}\hline
ID & Name & Age\\\hline

960114 & Yang & 20 \\\hline

\end{tabular}
}

\put(20,10){Table. 4: Result of example 3.}

\end{picture}

\subsection{Factorized quantum search algorithm} In \cite{Patel01},
Patel has observed that sorting can be thought of as factorization
of the search process and the location of a desired item in a sorted
database can be found by classical queries that inspect one letter
of the label at a time. Consider a collection of items in a
database, characterized by some property with values on the real
line. Sorting these items means arranging them in an ordered
sequence according to their property values. In order to efficiently
implement the sorting algorithm we need to digitize the property
values. It means replacing the property values by integer labels and
writing them as a string of letters belonging to a finite alphabet.
In digital computers, this finite alphabet has size 2, and the
letters are called bits \cite{Patel01}. Assume that the database
have $N=2^{n}$ items (If N is not a power of 2, then the database is
padded up with extra labels to make $N=2^{n}$.) and without loss of
generality the desired item has the label
\begin{equation}
x\equiv x_{1}x_{2}\ldots x_{n}=00\ldots 0.
\end{equation}
Then the search process is equivalent to finding x such that
\begin{equation}
f(x)=\prod_{i}f_{i}(x_{i})=(1-x_{1})(1-x_{2})\ldots (1-x_{n})
\end{equation}
equals to one. In the sorted and digitized database, one searches
for an item by inspecting only one $x_{i}$ at a time, i.e.,
evaluating f(x) by sequentially combining its factors $f_{i}(x_{i})$
. The functions $f(x)$ and $f_{i}(x_{i})$ are referred to as oracle
and factorized oracle respectively. If only the oracle $f(x)$ is
available then the search process requires $O(N/2)$ queries even
with a sorted database. The collection of factorized oracles
$f_{i}(x_{i})$ is more powerful than the global oracle $f(x)$ (One
can construct f(x) by combining all the $f_{i}(x_{i})$ together.).
And with the factorized oracles, the search process requires only
$O(lnN)$ queries \cite{Patel01}.

In the paper, Patel has also observed that if the database permits
quantum queries , then mere digitization is sufficient for efficient
search with the quantum factorized oracles. The algorithm called
factorized quantum search algorithm introduced by him is described
as
\begin{equation}
|x\rangle=\prod_{i=1}^{n}(P_{i}R_{i}F_{i})|\psi_{start}\rangle,
\end{equation}
where
$|\psi_{start}\rangle=\sum_{j=1}^{n}\frac{1}{\sqrt{N}}|j\rangle$,
which is the uniform superposition of all the N states. The
transformations $F_{i}$, $R_{i}$, and $P_{i}$ act on a single letter
of the label only. $F_{i}$ will evaluate to -1 when the letter and
its corresponding letter of the desired item match and to +1 when
they do not.
\begin{equation}
F_{i}=I_{1}\otimes \ldots \otimes \{\pm 1\}_{i} \otimes\ldots
\otimes I_{n}
\end{equation}
The reflection operator $R_{i}$ can also written in a factorized
form,
\begin{equation}
R_{i}=I_{1}\otimes \ldots \otimes \{R_{0}\}_{i} \otimes\ldots
\otimes I_{n},
\end{equation}
where
\begin{equation}
R_{0}=\frac{1}{2}
\begin{pmatrix}
-1& +1& +1& +1\\
+1& -1& +1& +1\\
+1& +1& -1& +1\\
+1& +1& +1& -1
\end{pmatrix}.
\end{equation}
The projection/measurement operator $P_{i}$ removes from the Hilbert
space all the states with zero amplitude. Note that the quantum
database is digitized with the size of the alphabet equals to 4. The
factorized quantum search algorithm locates the desired item in an
unsorted database using $O(log_{4}N)$ queries.

The search process inspects each letter of the label in turn, and
decide whether it matches the corresponding letter of the desired
string or not. The algorithm use Grover's algorithm
\cite{Grover96,Grover97} to do this job. Since Grover's algorithm
requires only one query to pick one item out of four with certainty,
it is efficient to digitize the quantum database with the alphabet
$\{|00\rangle,|01\rangle,|10\rangle,|11\rangle\}$. Digitization
requires that the items must be distinct from each other, and before
digitization, we might need to pad up the database. We will turn
back to these two problems later.

\section{The quantum search engine}
\subsection{The function of the engine}
The database we are going to searching for desired items is
unsorted. It is the only difference between the database and a
general database. Our quantum search engine on such a database can
be demonstrated by FIG.1, in which the Implementation module in
charge of carrying out an actual quantum database search algorithm
(i.e. Factorized quantum search algorithm here). The purpose of the
analysis module is to resolve the query conditions passed into the
engine by outside programs such as database application programs,
and initiates a sequence of calls to the implementation module.

\begin{picture}(200,150)
\put(5,30){\includegraphics{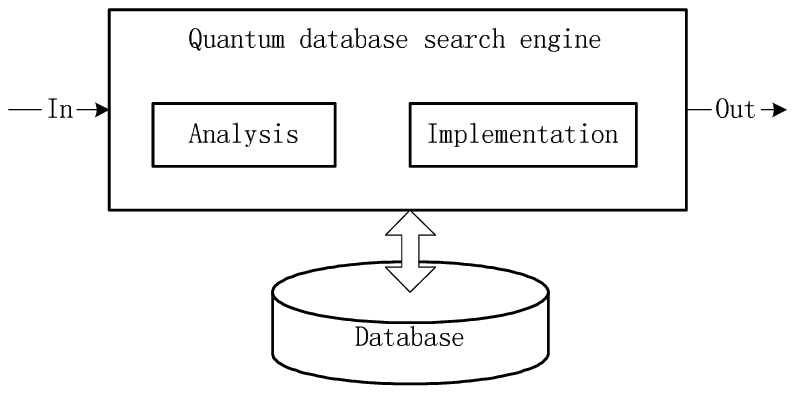}}

\put(5,15){FIG. 1: The function map of quantum search engine}
\end{picture}

We will return to the implementation module later, but in the rest
of this subsection, we focus on the analysis module. From the
previous section we can see that mere digitalization is sufficient
for the factorized quantum search algorithm. The digitalization
requires that the property values are distinct from each other. It
means that we can not apply the factorized quantum search algorithm
directly if there are more than one potentially desired items in the
query conditions. For example, the query indicated by the example 1
in section II can not carried out directly by implementing the
factorized quantum search algorithm, because query condition of it
includes 2 potential desired records. We have used the fact that if
one implements a query on a database management interface(Here, one
query means a SQL statement with a query condition, which is
different from the "query" used in query complexity theory), the
corresponding query request will be sent to the search engine.

The problem that how to digitize a database becomes how many
potential target items in one query. To answer this problem we need
to make query request clearer.

\begin{Defi}
A complex query is a SQL statement with a relational expression.
\end{Defi}

\begin{Defi}
A simple query is a SQL statement with a logical expression.
\end{Defi}

A query condition is equivalent to a relational expression, which
combines two or more logical expressions with relational operators
such as AND, OR and NOT. A logical expression is an expression that
has two operands connected with a logical operator from the set
$\{>, \geqslant, <, \leqslant, =, \neq\}$ \cite{Younes07}.

\begin{theorem}
Any complex query can be split into a sequence of simple queries.
\end{theorem}
To prove this theorem we need three facts:

Fact 1: Assume that a query condition has the form of "CONDITION1
AND CONDITION2" . We can divided it into two queries using
"CONDITION1" and "CONDITION2" respectively. They return with 2 set
of records. The intersection of these two sets gives the result of
the primitive query.

For example, the query demonstrated by example 3 can be divided into
two queries as indicated by example 1 and example 2 respectively.

Fact 2: Assume that a query condition has the form of "CONDITION1 OR
CONDITION2" . Similarly, the query can be separated into two
queries, and the union of the set returned by them gives the result
of the primitive query.

Fact 3: Assume that a query condition has the form of "NOT
CONDITION". The result of the primitive query is the complementary
set of the resulting set of records of the query with query
condition "CONDITION".

With these three facts, Theorem 1 is obvious. Here, we only show how
to split a complex query into simple query without considering the
problem of query optimization. Then, how many potential target
records in a simple query? It depends on both the content of the
database we are manipulating and the query itself. In the next two
subsections, we will consider how the effect of the content of the
database and how a simple query can be converted into a sequence of
calls to the factorized quantum search algorithm. But here, we
return to summarize the function of the search engine. It can be
described as follows:

a. Accept query request.

b. Call the analysis module to analyze the query condition.

c. Split the primitive query into simply queries.

d. Initiate a sequence of calls to the implementation module.

e. Collect the results returned by the calls and do further
operations.

f. Return the final result.

In short, the quantum search engine will convert every complex query
into a sequence of simple queries each with one logical expression,
and finally into a sequence of calls to the factorized quantum
search algorithm.

\subsection{Digitalization}
Digitization means writing all the property values as strings of
letters belonging to a finite alphabet \cite{Patel01}. This requires
that all the property values, among which the algorithm is going to
searching for desired items, must be distinct from each other.
Digitization on the key values is easy, because the key values are
distinct. Consider the case that if the content of TABLE 1 is the
content of the database we are going to query for target records. We
can use a simple coding scheme for this case by replacing $960112
\sim 960115$ by $0 \sim 3$, and digitize them as
\begin{equation}
\begin{split}
&0\longrightarrow |00\rangle\\
&1\longrightarrow |01\rangle\\
&2\longrightarrow |10\rangle\\
&3\longrightarrow |11\rangle.
\end{split}
\end{equation}

Digitization also requires that a database file must have $N=4^{n}$
items. If its original file does not meet this requirement, we have
to pad up the database file. And we need a tag to identify whether
or not a database file has been padded up. If implementation of
factorized quantum search algorithm gives a record with tag
indicating that it is a padding item, then the result would be
abandoned. But for simplicity, we assume the requirement is
satisfied (In fact, we can avoid this problem by allocating a
database file with size $L*4^{n}$, where L is the size of an item).

The values of common property are generally not distinct from each
other in an actual database. A solution to this problem is that
using one or more auxiliary files to record the values of common
property if we permit a search on the property. If a common property
has two or more auxiliary files (If the property values are distinct
from each other, then it doesn't need any auxiliary file), we say
that the database enable queries on the property. In other words, a
search on the values of a common property without the help of
auxiliary file will possibly fail. The auxiliary file has such a
structure as

\begin{picture}(180,50)
\put(20,30){
\begin{tabular}{|c |c |}\hline
Property values & Addresses in main file\\\hline
\end{tabular}
}

\put(24,10){Table. 5: Structure of an auxiliary file.}
\end{picture}

A table in a database corresponds to a main file and two or more
auxiliary files. If the database detects that a manipulation on it
will result in non-uniqueness of one or more property values, it
will record the property value in an auxiliary file. It needs one or
more auxiliary files for a property. The number of the auxiliary
files corresponding to a property rests with how many property
values are the same.

For example, to record the property values of "Age", it might need
such two auxiliary files as

\begin{picture}(200,180)
\put(20,140){
\begin{tabular}{|c |c |}\hline
Property values & Addresses in main file\\\hline

18 & 1\\\hline

20 & 2\\\hline

19 & 3\\\hline
\end{tabular}
}

\put(80,100){Table. 6, A}

\put(20,60){
\begin{tabular}{|c |c |}\hline
Property values & Addresses in main file\\\hline

20 & 4\\\hline
\end{tabular}
}

\put(80,30){Table. 6, B}

 \put(24,10){Table. 6: Auxiliary files for table 1.}
\end{picture}

Here, we suppose the addresses of the four records in table 1 are 1,
2, 3, and 4. Note that all the property values in an auxiliary file
should be distinct from each other. So we can treat the property
values in a auxiliary file in the way as the key values when
digitalizing.

\subsection{Initiation of calls to implementation module}
In this subsection, we focus on a simple query with only a logical
expression as its query condition. We have said that a logical
expression is an expression that has two operands connected with a
logical operator from the set $\{>, \geqslant, <, \leqslant, =,
\neq\}$ \cite{Younes07}. A logical expression in a simply query has
such a form as "x o a", where x is a property variable, o is a
logical operator, and a is a given property value. Assume that the
definition domain of x is $[x_{1},x_{2}]$. The algorithm of
initiating calls to implementation module can be described as the
following subroutines each corresponding to a type of logical
expression.

\textbf{f(x,=,a):}

a. If x is a key property, then call the implementation module with
the query condition on the main file.

b. Else if the search on the property x is not allowable, i.e., its
property values are not distinct from each other and no auxiliary
file are available, then return with failure.

c. For Every auxiliary file, call the implementation module with the
query condition.

d. Return with the union set of the results of all calls.

\textbf{f(x,$\neq$,a):}

a. For every y in $[x_{1},x_{2}]$ and unequal to a, call f(x,=,y).

b. Return with the union set of the results of all calls.

\textbf{f(x,$<$,a):}

a. For Every y in $[x_{1},a)$, call f(x,=,y).

b. Return with the union set of the result of all calls.

The subroutines for the rest types of logical expression, i.e.,
f(x,$\leqslant$,a), f(x,$>$,a), and f(x,$\geqslant$,a), are similar
to f(x,$<$,a).

\subsection{The implementation module}

We have known that the implementation module carrys out the
factorized quantum search algorithm. So, we turn to see how the
factorized quantum search algorithm work on an actual quantum
computer. Assume that our database exists in a qRAM, the content of
the database is table 1, and we want to find the record that
"ID=960112". The four property values of "ID" are encoded as

\begin{equation}
\begin{split}
&960112\longrightarrow |00\rangle\\
&960113\longrightarrow |01\rangle\\
&960114\longrightarrow |10\rangle\\
&960115\longrightarrow |11\rangle.
\end{split}
\end{equation}

First, we need to prepare the content of the database in a uniform
state. we can do this job by setting the address register A in the
state $\frac{1}{2}(|a\rangle+|b\rangle+|c\rangle+|d\rangle)$, where
a, b, c, d, are the corresponding addresses of the four records. And
the qRAM will return a superposition of data in a data register D,
correlated with the address register
\begin{equation}
\frac{1}{2}(|00\rangle\otimes|a\rangle+|01\rangle\otimes|b\rangle+|10\rangle\otimes|c\rangle+|11\rangle\otimes|d\rangle)
\end{equation}

It is easy to see that it takes only one oracle query for the
factorized quantum search algorithm to find the desired item, and
the resulting state of the register would be
$|00\rangle\otimes|a\rangle$. Measurement on the register gives the
desired item with certainty.

\subsection{query complexity and space consumption}

If a property has t possible values, and $j_{i}$ denotes the number
of items with the property value $p_{i}$, where i ranges from 1 to
t, then the number of the auxiliary files is $M=max(j_{1}, j_{2},
\ldots , j_{t})$. Let the size of the M auxiliary files be $N_{1},
N_{2},\ldots, N_{M}$, then the total memory space for all the M
auxiliary files is $N_{1}+N_{2}+\ldots +N_{M}$, which is $4N$ in the
worst case (Note that we need to pad every auxiliary file such that
$N_{i}$ is the power of 4).

\begin{theorem}
The query complexity of Our algorithm for the quantum search engine
responding to one complex query request is at most
$O(P*Q*M*log_{4}N)$, where P is the number of the potential simple
query requests in the complex query request, Q is the maximum number
of calls to the factorized quantum search algorithm, M is the number
of the auxiliary files for the property on which our algorithm are
searching for desired items.
\end{theorem}
Proof: Because a complex query is converted into $P*Q$ calls to the
factorized quantum search algorithm on every auxiliary files, the
total number of calls to the factorized quantum search algorithm is
$P*Q*M$. And the query complexity of the factorized quantum search
algorithm is $O(log_{4}N)$ \cite{Patel01}. So, the total query
complexity corresponding to a complex query is at most
$O(P*Q*M*log_{4}N)$.

If the values of the property, on which our algorithm are
implementing, are distinct from each other, then the query
complexity are $O(P*Q*log_{4}N)$. However, the query complexity of a
typical search engine using the best classical search algorithm
mentioned above is $O(P*Q*log_{2}N)$ for a complex query.

\section{Examples}
We take the example 3 in section II as an example to show how our
quantum search engine works. After the engine accept the query
request, it will first call the analysis module to analyze the query
condition, and split the primitive query into simply queries
demonstrated by example 1 and 2. Let us see how the engine deals
with the example 1 and 2. Assume that "ID" is the key property and
the database enable queries on the property "Age".

\textbf{Search process of example 1:}

a. The engine calls the subroutine f(Age,=,20).

b. Because "Age" is not the key property, then for every auxiliary
files(The auxiliary files can be seen in table 6.), call the
implementation module with the query condition.

c. For the first auxiliary file, the implementation module returns
with

\begin{picture}(120,50)
\put(40,20){
\begin{tabular}{|c |c |c|}\hline
ID & Name & Age\\\hline

960114 & Yang & 20\\\hline

\end{tabular}
}

\end{picture}

For the second auxiliary file, the implementation module returns
with

\begin{picture}(120,50)
\put(40,20){
\begin{tabular}{|c |c |c|}\hline
ID & Name & Age\\\hline

 960115 & Yingyu & 20\\\hline

\end{tabular}
}

\end{picture}

d. The result of example 1, which can seen in table 2, is the union
of these two results.

\textbf{Search process of example 2:}

a. The engine calls the subroutine f(ID,$<$,960115).

b. The subroutine then call f(ID,=,960112), f(ID,=,960113), and
f(ID,=,960114). Note that ID is a key property, then f(ID,=,960112)
and f(ID,=,960113) call the implementation module with the
corresponding query condition respectively on the main file.

c. f(ID,=,960112) returns with

\begin{picture}(120,50)
\put(40,20){
\begin{tabular}{|c |c |c |}\hline
ID & Name & Age\\\hline

960112 & Lin & 18 \\\hline

\end{tabular}
}

\end{picture}

f(A,=,960113) returns with

\begin{picture}(120,50)
\put(40,20){
\begin{tabular}{|c |c |c |}\hline
ID & Name & Age\\\hline

960113 & Lin & 19 \\\hline

\end{tabular}
}

\end{picture}

f(A,=,960114) returns with

\begin{picture}(120,50)
\put(40,20){
\begin{tabular}{|c |c |c |}\hline
ID & Name & Age\\\hline

960114 & Lin & 20 \\\hline

\end{tabular}
}

\end{picture}

The union of the results of f(ID,=,960112), f(A,=,960113), and
f(A,=,960114) gives the results of example 2, which can be seen in
table 3. And the intersection of the results of example 1 and 2 is
the result of example 3, which can be seen in table 4.
\section{Conclusion}
We have considered the problem of finding one or more desired items
out of an unsorted database. The factorized quantum search algorithm
presented by Patel can locate one desired item in an unsorted
database using O(log4N) queries to factorized oracles. But the
algorithm can only solve the problem finding the desired item out of
an unsorted database with only one target item. We extend the
algorithm to solve the problem with more than one target items in an
unsorted database. The goal is achieved by introducing auxiliary
file, and converting a complex query request into a sequence of
simple queries, and finally into a sequence of calls to the
factorized quantum search algorithm.

The query complexity of our algorithm is $O(P*Q*M*log_{4}N)$, where
P is the number of the potential simple query requests in the
complex query request, Q is the maximum number of calls to the
factorized quantum search algorithm of the simple queries, M is the
number of the auxiliary files for the property on which our
algorithm are searching for desired items. We know that the query
complexity of the best classical algorithm to select P*Q items from
a sorted database is $P*Q*log_{2}N$. When compared with the
classical algorithm on either sorted or unsorted database, if M is
small, the quantum search engine on unsorted database presented in
the paper is more efficient.

The most important thing implied in this paper is that to manage an
unsorted database on a quantum computer is possible and efficient.
Though a query algorithm on a sorted database can be more efficient
than a query algorithm on an unsorted database, the sorting price
might be very high. So the quantum search algorithm presented in
this paper offers another way to database management on an actual
quantum computer.


\begin{thebibliography}{10}
\expandafter\ifx\csname
natexlab\endcsname\relax\def\natexlab#1{#1}\fi
\expandafter\ifx\csname bibnamefont\endcsname\relax
  \def\bibnamefont#1{#1}\fi
\expandafter\ifx\csname bibfnamefont\endcsname\relax
  \def\bibfnamefont#1{#1}\fi
\expandafter\ifx\csname citenamefont\endcsname\relax
  \def\citenamefont#1{#1}\fi
\expandafter\ifx\csname url\endcsname\relax
  \def\url#1{\texttt{#1}}\fi
\expandafter\ifx\csname urlprefix\endcsname\relax\def\urlprefix{URL
}\fi \providecommand{\bibinfo}[2]{#2}
\providecommand{\eprint}[2][]{\url{#2}}

\bibitem[{\citenamefont{Patel}(2001)}]{Patel01}
\bibinfo{author}{\bibfnamefont{A.}~\bibnamefont{Patel}},
  \bibinfo{journal}{Phys. Rev. A} \textbf{\bibinfo{volume}{64}},
  \bibinfo{pages}{034303} (\bibinfo{year}{2001}).

\bibitem[{\citenamefont{Hoyer and Neerbek}(2000)}]{Hoyer00}
\bibinfo{author}{\bibfnamefont{P.}~\bibnamefont{Hoyer}} \bibnamefont{and}
  \bibinfo{author}{\bibfnamefont{J.}~\bibnamefont{Neerbek}},
  \bibinfo{journal}{arXiv:quant-ph/0009032}  (\bibinfo{year}{2000}).

\bibitem[{\citenamefont{Farhi et~al.}(1999)\citenamefont{Farhi, Goldstone,
  Gutmann, and Sipser}}]{Farhi99}
\bibinfo{author}{\bibfnamefont{E.}~\bibnamefont{Farhi}},
  \bibinfo{author}{\bibfnamefont{J.}~\bibnamefont{Goldstone}},
  \bibinfo{author}{\bibfnamefont{S.}~\bibnamefont{Gutmann}}, \bibnamefont{and}
  \bibinfo{author}{\bibfnamefont{M.}~\bibnamefont{Sipser}},
  \bibinfo{journal}{arXiv:quant-ph/0009032}  (\bibinfo{year}{1999}).

\bibitem[{\citenamefont{Childs et~al.}(2007)\citenamefont{Childs, Landahl, and
  Parrilo}}]{Childs07}
\bibinfo{author}{\bibfnamefont{A.~M.} \bibnamefont{Childs}},
  \bibinfo{author}{\bibfnamefont{A.~J.} \bibnamefont{Landahl}},
  \bibnamefont{and} \bibinfo{author}{\bibfnamefont{P.~A.}
  \bibnamefont{Parrilo}}, \bibinfo{journal}{Phys. Rev. A}
  \textbf{\bibinfo{volume}{75}}, \bibinfo{pages}{032335}
  (\bibinfo{year}{2007}).

\bibitem[{\citenamefont{Nielson and Chuang}(2000)}]{Nielson00}
\bibinfo{author}{\bibfnamefont{M.~A.} \bibnamefont{Nielson}} \bibnamefont{and}
  \bibinfo{author}{\bibfnamefont{I.~L.} \bibnamefont{Chuang}},
  \emph{\bibinfo{title}{Quantum Computation and Quantum Information}}
  (\bibinfo{publisher}{Cambridge University Press},
  \bibinfo{address}{Cambridge}, \bibinfo{year}{2000}).

\bibitem[{\citenamefont{Giovannetti et~al.}(2008)\citenamefont{Giovannetti,
  Lloyd, and maccone}}]{Giovannetti08}
\bibinfo{author}{\bibfnamefont{V.}~\bibnamefont{Giovannetti}},
  \bibinfo{author}{\bibfnamefont{S.}~\bibnamefont{Lloyd}}, \bibnamefont{and}
  \bibinfo{author}{\bibfnamefont{L.}~\bibnamefont{maccone}},
  \bibinfo{journal}{Phys. Rev. Lett} \textbf{\bibinfo{volume}{100}},
  \bibinfo{pages}{160501} (\bibinfo{year}{2008}).

\bibitem[{\citenamefont{Elmasri and Navathe}(2006)}]{Elmasri06}
\bibinfo{author}{\bibfnamefont{R.~R.} \bibnamefont{Elmasri}} \bibnamefont{and}
  \bibinfo{author}{\bibfnamefont{S.~B.} \bibnamefont{Navathe}},
  \emph{\bibinfo{title}{Fundamentals of Database System}}
  (\bibinfo{publisher}{Addison Wesley}, \bibinfo{address}{Boston},
  \bibinfo{year}{2006}).

\bibitem[{\citenamefont{Younes}(2007)}]{Younes07}
\bibinfo{author}{\bibfnamefont{A.}~\bibnamefont{Younes}},
  \bibinfo{journal}{arXiv e-Print quant-ph/0705.4303}  (\bibinfo{year}{2007}).

\bibitem[{\citenamefont{Grover}(1996)}]{Grover96}
\bibinfo{author}{\bibfnamefont{L.~K.} \bibnamefont{Grover}}, in
  \emph{\bibinfo{booktitle}{Proceedings of 28th Annual ACM Symposium on Theory
  of Computing}} (\bibinfo{year}{1996}), pp. \bibinfo{pages}{212--219}.

\bibitem[{\citenamefont{Grover}(1997)}]{Grover97}
\bibinfo{author}{\bibfnamefont{L.~K.} \bibnamefont{Grover}},
  \bibinfo{journal}{Phys. Rev. Lett.} \textbf{\bibinfo{volume}{79}},
  \bibinfo{pages}{325} (\bibinfo{year}{1997}).

\end{thebibliography}
\end{document}